\documentclass[preprintnumbers,article,amsmath,amssymb,floatfix,10pt,prd,onecolumn,superscriptaddress,nofootinbib]{revtex4-2}
\usepackage{bm}
\usepackage{amsfonts}
\usepackage{latexsym}
\usepackage[latin1]{inputenc}
\usepackage{graphicx}
\usepackage{amsmath}
\usepackage{palatino}
\usepackage{mathpazo}
\usepackage{textcomp}
\usepackage{textcomp}
\usepackage{mathrsfs}
\usepackage{float}
\usepackage{booktabs}
\usepackage{dcolumn}
\usepackage{ragged2e}
\usepackage{hyperref}
\usepackage{tensor}
\usepackage[version=3]{mhchem}
\hypersetup{colorlinks,citecolor=blue}
\hypersetup{colorlinks=true,linkcolor=red,filecolor=magenta,    urlcolor=blue}
\usepackage{amsmath}
\usepackage{xcolor}
\usepackage{enumitem}
\usepackage{orcidlink}
\usepackage{epsfig}
\usepackage{subfigure}
\usepackage{commath}
\usepackage{caption}

\begin{document}

\title{BBN to Late-Time Acceleration in $f(T,\mathcal{L}_m)$ Gravity}
\author{Sai Swagat Mishra\orcidlink{0000-0003-0580-0798}}
\email{saiswagat009@gmail.com}
\affiliation{Department of Mathematics, Birla Institute of Technology and Science, Pilani, Hyderabad Campus, Jawahar Nagar, Kapra Mandal, Medchal District, Telangana 500078, India}%

\author{Suchita Patel\orcidlink{0009-0003-9597-7438}}
\email{kmsuchitapatel017@gmail.com}
\affiliation{Department of Mathematics, Birla Institute of Technology and Science, Pilani, Hyderabad Campus, Jawahar Nagar, Kapra Mandal, Medchal District, Telangana 500078, India}%

\author{P.K. Sahoo\orcidlink{0000-0003-2130-8832}}
\email{pksahoo@hyderabad.bits-pilani.ac.in}
\affiliation{Department of Mathematics, Birla Institute of Technology and Science, Pilani, Hyderabad Campus, Jawahar Nagar, Kapra Mandal, Medchal District, Telangana 500078, India}
\begin{abstract}
    We present, to our knowledge, the first systematic study of early-late cosmic evolution and acceleration in the framework of $f(T,\mathcal{L}_m)$ gravity, an extension of teleparallel theories coupling torsion with the matter Lagrangian. By incorporating the Big-Bang Nucleosynthesis (BBN) bound on the freeze-out temperature, we obtain a tight constraint on the inverse-torsion parameter, ensuring consistency with early-time physics.
 Employing Markov Chain Monte Carlo analyses with progressively richer observational datasets, CC, Union3, and SN22 supernovae, we constrain a well-motivated model and reconstruct key cosmological functions. The reconstructed Hubble and distance modulus functions show excellent agreement with the observations, confirming the observational viability of the model.
    The model successfully reproduces the observed late-time expansion history, yielding a transition from deceleration to acceleration through the deceleration parameter. The effective equation of state is found to remain negative throughout, with present values $w_0 > -1$, indicating a quintessence-like behavior rather than a cosmological constant or phantom regime. These results highlight the ability of $f(T,\mathcal{L}_m)$ gravity to mimic the concordance scenario while allowing controlled deviations in the expansion history.  
\end{abstract}
\maketitle
\section{Introduction}
In the quest to understand the observed Universe, theory is still striving to keep pace with the precision of modern data. From the role of dark matter in the growth of cosmic structures to the large-scale expansion history, no single framework provides a complete explanation. A striking example is the late-time accelerated expansion of the Universe, first established through Type Ia supernova observations by Riess et al. and Perlmutter et al. at the end of the 1990s \cite{SupernovaSearchTeam:1998fmf, SupernovaCosmologyProject:1997czu, SupernovaCosmologyProject:1998vns, SDSS:2005xqv, DESI:2024mwx}. While the standard approach attributes this phenomenon to a cosmological constant within the Lambda Cold Dark Matter ($\Lambda$CDM) model, it is well understood that Einstein's General Relativity (GR) requires extensions at large scales, and that it must ultimately be reconciled with quantum mechanics to describe the high-energy conditions of the early Universe. Over the years, numerous proposals have attempted to bridge these gaps, yet none has emerged as a complete and definitive theory.

The $\Lambda$CDM model has nevertheless been extraordinarily successful in explaining a wide range of cosmological observations, from the cosmic microwave background to baryon acoustic oscillations and supernovae\cite{Rubin:2023jdq, Jimenez:2001gg, Jimenez:2003iv}. However, it suffers from well-known theoretical challenges, such as the cosmological constant problem \cite{Weinberg:1988cp} and the coincidence problem \cite{Steinhardt1999}. On top of these conceptual issues, increasingly precise datasets have revealed persistent observational tensions, including the discrepancy between early- and late-time determinations of the Hubble constant $H_0$ \cite{Ivanov:2020mfr}, and differences in the inferred clustering amplitude $S_8$ \cite{Planck:2018vyg, Verde2019, DiValentino2021, Mandal:2023bzo}. These challenges have motivated the development of alternative scenarios in which cosmic acceleration emerges from modifications to the underlying theory of gravity.

In recent years, geometrical modifications of GR have achieved significant success in addressing cosmological and astrophysical phenomena beyond the scope of the standard framework. These theories extend or reformulate the underlying geometrical structure of gravity, offering new mechanisms to explain late-time cosmic acceleration, the nature of dark energy, and other outstanding puzzles \cite{Heisenberg:2023lru,Bahamonde:2021gfp, Shankaranarayanan:2022wbx, Clifton:2011jh, Mishra:2024oln}. One particularly appealing direction is offered by teleparallel gravity, where torsion rather than curvature is used to describe gravitational interactions \cite{Aldrovandi:2013wha}. The teleparallel equivalent of GR (TEGR) reproduces Einstein's equations at the background level, while extensions such as $f(T)$ gravity, where $T$ is the torsion scalar, can naturally drive late-time acceleration without invoking a cosmological constant \cite{Cai:2015emx}. A nonminimal matter coupling torsion was first performed by Harko et al. \cite{Harko:2014sja}, where the authors considered a specific class of the coupling. Recently, Cruz et al. \cite{Cruz:2025fuk} considered a more general coupling functional form of $f(T,\mathcal{L}_m)$ to investigate the gravitational baryogenesis scenario.
This nonminimal coupling between geometry and matter introduces a richer phenomenology, opening the door to effective interactions between dark energy and matter sectors.

In addition to late-time probes, early-Universe physics provides powerful
constraints on modified gravity scenarios. In particular, Big Bang
Nucleosynthesis (BBN) is highly sensitive to deviations in the expansion rate
during the radiation era, allowing one to impose stringent bounds on any
modification that affects $H(z)$ at early times \cite{Steigman:2007xt, Capozziello:2017bxm, Mishra:2023onl, Kavya:2024ssu}. 
In the present work, we incorporate the BBN freeze-out temperature constraint as
an independent and complementary test of the $f(T,\mathcal{L}_m)$ framework.

Despite its theoretical promise, the cosmological implications of
$f(T,\mathcal{L}_m)$ gravity remain largely unexplored. In particular, its
compatibility with both early-time constraints, such as those from BBN, and with the observed late-time expansion history has not been
systematically tested. In this work, we aim to fill this gap by presenting, to
our knowledge, the first comprehensive early-late cosmological analysis of the
$f(T,\mathcal{L}_m)$ framework. Using Markov Chain Monte Carlo (MCMC) methods with a
hierarchy of recent datasets, we constrain the theory and assess its potential
departures from $\Lambda$CDM, thereby providing new insights into the role of
torsion-matter couplings in shaping the evolution of the Universe.

The paper is organized as follows. In Sec.~\ref{sec:back}, we introduce 
the $f(T,\mathcal{L}_m)$ framework and summarize the background cosmological 
equations relevant for our analysis. In Sec.~\ref{sec:bbn}, we present the 
BBN formalism, and in Sec.~\ref{sec:model} we specify the inverse-torsion 
$f(T,\mathcal{L}_m)$ model considered in this work and derive the expressions 
governing both the early- and late-time cosmological evolution. In 
Sec.~\ref{sec:data}, we describe the observational datasets employed, including 
cosmic chronometers, Union3, and SN22 supernovae. The main results, including 
the BBN bounds, reconstructed cosmological functions, and parameter constraints, 
are presented in Sec.~\ref{sec:results}. Finally, in 
Sec.~\ref{sec:conclusions}, we summarize our findings and discuss their 
implications for cosmic acceleration and possible future directions.

\section{Background Equations of $f(T,\mathcal{L}_m)$ gravity}\label{sec:back}
In this section, we present the framework of $f(T,\mathcal{L}_m)$ gravity and derive the fundamental governing the background cosmological dynamics. 
In teleparallel gravity, which is the theory that adopts Weitzenb\"{o}ck connection without considering the Levi-Civita connection, tetrads are the key fields instead of the metric as in general relativity. These tetrads, denoted by $e^{\mu}_{\ \ i}$ express the dynamic field of the theory and satisfy the following orthonormality condition
\begin{equation}\label{tetrad}
    e^{\ \alpha}_{i}e^{\ \beta}_{j}g_{\alpha\beta}=\eta_{ij},
\end{equation}
where  $g_{\alpha\beta}$ is the metric of the curved space time and $\eta_{ij}$ is the flat space metric tensor $\eta_{ij}=diag(1,-1,-1,-1)$.
Since $e^{\alpha}_{\ \ i} e_{\beta}^{\ \ i}=\delta^{\alpha}_{\ \ \beta}$, therefore equation (\ref{tetrad}) can also be written as 
\begin{equation}
    e^{i}_{\  \alpha}e^j_{\  \beta}\eta_{ij}=g_{\alpha\beta}.
\end{equation}
The Weitzenb\"{o}ck's connection \cite{Aldrovandi:2013wha} in terms of tetrads is defined as
\begin{equation}
    \Gamma^{\rho}_{\alpha\beta}=e_i^{\ \rho}\partial_{\beta}e^{\ \ i}_{\alpha}.
\end{equation}
Through the antisymmetric part of the above connection, the torsion tensor is defined as
\begin{equation}\label{Torsion}
    T^{\lambda}_{\ \ \alpha\beta}=\Gamma^{\lambda}_{\alpha\beta}-\Gamma^{\lambda}_{\beta\alpha}=e_i^{\ \ \lambda}(\partial_{\alpha}e^i_{\ \ \beta}-\partial_{\beta}e^i_{\ \ \alpha}).
\end{equation}
The contorsion tensor is defined as
\begin{equation}
    K^{\alpha\beta}_{ \ \  \  \  \lambda}=\frac{-1}{2}(T^{\alpha\beta}_{\ \  \  \  \lambda}-T^{\beta\alpha}_{\ \  \  \  \lambda}-T_{\lambda}^{\ \ \alpha\beta}).
\end{equation}
And the superpotential tensor, denoted as
\begin{equation}
    S_{\lambda}^{\ \  \alpha\beta}=\frac{1}{2}(K^{\alpha\beta}_{ \ \  \  \  \lambda}+\delta_{\lambda}^{\alpha}T^{\mu\beta}_{\ \ \mu}-\delta_{\lambda}^{\beta}T^{\mu\alpha}_{\ \ \mu}).
\end{equation}
Now, the torsion scalar is given by the following contraction
\begin{equation}\label{torsion scalar}
    T= T^{\lambda}_{\  \ \alpha\beta}S_{\lambda}^{\ \  \alpha\beta}.
\end{equation}
Finally, the action for $f(T,\mathcal{L}_m)$ is given as \cite{Cruz:2025fuk}
\begin{equation}\label{action}
    \mathcal{S}=\frac{\int dx^4 e[T+f(T,\mathcal{L}_m)]}{16\pi G}+\int dx^4 e\mathcal{L}_m  ,
\end{equation}
Note that $e=det(e^{i}_{\ \alpha})$. Here $\mathcal{L}_m$ is the Lagrangian matter field, with the assumption that it depends only on the tetrads but not on their derivatives, and G is the Newtonian constant.
The variation of the action ( \ref{action}) with respect to tetrads gives rise to the following field equations,
\begin{equation}
    \begin{split}
        &(1+f_T)\left[e^{-1}\partial_\alpha(ee_i^{\ \lambda}S_\lambda^{\ \alpha\beta})-e_i^{\ \rho}T^{\lambda}_{\ \alpha\rho}S_\lambda^{\ \beta\alpha}\right]+e_i^{\ \lambda}S_\lambda^{\ \alpha\beta}(f_{TT}\partial_\alpha T+f_{T\mathcal{L}_m}\partial_\alpha \mathcal{L}_m)\\&-\frac{1}{4}f_{\mathcal{L}_m}\left(e_i^{\ \lambda}\overset{em}{T}{}_\lambda^{\ \beta}+e_i^{\ \lambda}\mathcal{L}_m\right)+e_i^{\ \beta}\left(\frac{T+f}{4}\right)=4\pi G e_i^{\ \lambda}\overset{em}{T}{}_\lambda^{\ \beta}\,,
    \end{split}
    \label{field equations}
\end{equation}
 where $\overset{em}{T}{}_\lambda^{\ \beta}$ represents the energy-momentum tensor of a perfect fluid (\cite{Aldrovandi:2013wha})
\begin{equation} 
 \frac{\delta e\mathcal{L}_m}{\delta e^i_{\ \beta}}=-ee_i^{\ \lambda}T_\lambda^{\ \beta},
 \end{equation} 
 and $f_T=\frac{\partial f}{\partial T}$, $f_{TT}=\frac{\partial^2 f}{\partial T^2}$, $f_{T{\mathcal{L}_m}}=\frac{\partial^2 f}{\partial T\partial{\mathcal{L}_m}}$, $f_{\mathcal{L}_m}=\frac{\partial f}{\partial{\mathcal{L}_m}}$.
We adopt the spatially flat FLRW metric, considering the spatial isotropy and homogeneity of the universe:
\begin{equation}\label{FLRW}
    ds^2=dt^2-a^2(t)((dx^1)^2+(dx^2)^2+(dx^3)^2),
\end{equation}
 where $a(t)$ is the scale factor as a function of time.
 Consequently, the suitable choice of tetrads for the metric (\ref{FLRW}) is the following diagonal form;
 \begin{equation}\label{tetrad diagonal form}
    e_i^{\ \alpha}=diag(1,a(t),a(t),a(t)).
 \end{equation}
 Substituting the tetrads from equation (\ref{tetrad diagonal form}) into the field equations \eqref{field equations}, we get the following modified Friedmann equations:
\begin{equation}
    3H^2=8\pi G\rho-6H^2 f_T+\frac{1}{2}f_{\mathcal{L}_m}\left(\rho+{\mathcal{L}_m}\right)-\frac{f}{2}\, ,
    \label{first Friedmann}
\end{equation}
\begin{equation}
        \Dot{H}=-4\pi G(\rho+p)-\Dot{H}(f_T-12H^2f_{TT})-Hf_{T{\mathcal{L}_m}}\partial_t{\mathcal{L}_m}-\frac{1}{2}f_{\mathcal{L}_m}\left(p+{\mathcal{L}_m}\right)\,,
\end{equation}
where $H$ is the Hubble parameter with $H=\frac{\dot{a}}{a}$. The pressure and energy density are given by $p$ and $\rho$, respectively. 

By using the equations (\ref{tetrad}), (\ref{Torsion}), and (\ref{torsion scalar}), the simplified torsion scalar in terms of the Hubble parameter for the FLRW metric is of the following form,
\begin{equation}
    T=-6H^2\,.
    \label{T}
\end{equation}

One can rewrite the above modified Friedmann equations in a GR equivalent form as
\begin{equation}
    3 H^2= 8\pi G\left(\rho+\rho_{DE}\right)\label{Hubble}\, ,
\end{equation}
\begin{equation}
    2\dot{H}+3H^2=-8\pi G \left(p+p_{DE}\right)\, ,
\end{equation}
where
\begin{equation}
    \rho_{DE}=\frac{3}{8\pi G}\left[-2H^2f_T+\frac{1}{6}f_{\mathcal{L}_m}\left(\rho+\mathcal{L}_m\right)-\frac{f}{6}\right]\label{rho_{DE}}\, ,
\end{equation}
and
\begin{equation}
       p_{DE}=-\rho_{DE}+\frac{1}{8\pi G}\Big[2\Dot{H}(f_T-12H^2f_{TT})+2H\partial_t(\mathcal{L}_m)f_{T{\mathcal{L}_m}}+f_{\mathcal{L}_m}(p+\mathcal{L}_m)\Big]\, .\label{Constrain1}
\end{equation}

To determine the continuity equation, we use the Lagrangian ${\mathcal{L}_m}=p$. Choosing instead ${\mathcal{L}_m}=-\rho$, leads to an additional force, which directly arises from the effect of the coupling mechanism. Nevertheless, even with ${\mathcal{L}_m}=p$, the persistence of the energy momentum source term indicates that the gravitational sector remains intrinsically non-conservative \cite{KavyaVenkatesha:2024bpj}.

To determine the deceleration parameter $q(z)$ in a model-independent way, one can employ the technique of cosmography, which relies on the expansion of the Taylor series for the scale factor $a(t)$ when $t=t_0$:
\begin{equation}
a(t) = a(t_0) \left[ 1 + H_0 (t-t_0) - \frac{1}{2} q_0 H_0^2 (t-t_0)^2 + \dots \right],
\end{equation}
where $H_0$ and $q_0$ denote the present-day values of the Hubble and deceleration parameters, respectively.

The deceleration parameter, in terms of time, can be defined as
\begin{equation}
q(t) = - \frac{1}{H^2} \frac{\ddot{a}}{a} = -1 - \frac{\dot{H}}{H^2}.
\end{equation}

Using the standard relation between the scale factor and redshift,
\begin{equation}
a(t) = \frac{1}{1+z} \quad \Rightarrow \quad \frac{d}{dt} = - (1+z) H(z) \frac{d}{dz},
\end{equation}
the deceleration parameter can also be rewritten as a function of redshift \cite{Capozziello:2014zda}
\begin{equation}
q(z) = -1 + \frac{(1+z)}{H(z)}\frac{d}{dz}(H(z)).
\end{equation}

The effective equation of state (EoS) parameter can
be defined as
\[
w_{eff}(z) = \frac{p_{eff}(z)}{\rho_{eff}(z)},
\]
where $p_{eff}(z)= p+p_{DE}$ and $\rho_{eff}(z)=\rho+\rho_{DE} $.
\section{BBN constraints on $ f(T,\mathcal{L}_m)$ gravity}\label{sec:bbn}

This section will develop the relationship between the BBN constraints and the deviation of the freeze-out time $\mathcal{T}_f$. BBN was observed throughout the radiation era \cite{Bernstein:1988ad, Dolgov:2002dc, Olive:1999ij}. In the context of standard GR, the first Friedmann equation can be roughly expressed as

\begin{equation}
    3H^2 = M_p^{-2} \rho,
\end{equation}
where \( M_p=\frac{1}{\sqrt{8\pi G}}=1.22 \times 10^{19} GeV\) is the reduced Planck mass. During the radiation-dominated era, the energy density is primarily due to relativistic particles, so

\begin{equation}\label{rho_r}
    \rho_r = \frac{\pi^2}{30} g_* \mathcal{T}^4,
\end{equation}
where \( g_* \) is the effective number of relativistic degrees of freedom usually approximated as \( g_* \sim 10\), and \( \mathcal{T}\) is the temperature. The Hubble parameter in this era is,
\begin{equation}\label{HGR} 
H^2 \approx \frac{1}{3M_p^2}\rho_r\equiv H^2_{GR}
\end{equation}
Therefore, the Hubble rate as a function of temperature is given by 
\begin{equation}\label{Hubble at temp}
    H(\mathcal{T}) \approx \left( \frac{\pi^2 g_*}{90} \right)^{1/2} \frac{\mathcal{T}^2}{M_p}.
\end{equation}
In the radiation era $a(t)\sim t^{\frac{1}{2}}$ and consequently $H(t)\sim \frac{1}{2t}$, this leads,
\begin{equation}
    \frac{1}{t} \approx \left( \frac{2 \pi^2 g_*}{45} \right)^{1/2} \frac{\mathcal{T}^2}{M_p}.
\end{equation}
The emergence of primordial $^4\mathrm{He}$ occurred during the initial phase of cosmic expansion, corresponding to a temperature of approximately $\mathcal{T} \sim 100~\mathrm{MeV}$. At this period, both the energy and particle densities of the Universe were primarily dictated by relativistic constituents, photons together with leptons such as electrons, positrons, and neutrinos. In contrast, the relatively minor presence of baryons, including neutrons and protons, contributed insignificantly to the overall energy density. All species within the primordial plasma remained thermally equilibrated owing to the high frequency of their mutual interactions. Moreover, protons and neutrons specifically preserved equilibrium through weak scattering processes involving leptons \cite{Lambiase:2005kb},
\begin{eqnarray}\label{weak interaction}
n + \nu_e \leftrightarrow p + e^-,\nonumber \\
n + e^+ \leftrightarrow p + \overline{\nu}_e,\\
n \leftrightarrow p + e^- + \overline{\nu}_e.\nonumber
\end{eqnarray}

During the BBN era, the determination of the neutron abundance is derived from the dynamics of the weak interaction processes governing proton-neutron conversion rate $\lambda_{pn}(\mathcal{T})$, and its inverse reaction $\lambda_{np}(\mathcal{T})$.
The total conversion rate \( \lambda_{\text{tot}}(\mathcal{T}) \) is the sum of these individual rates
\begin{equation}
\lambda_{\text{tot}}(\mathcal{T}) = \lambda(n + \nu_e \leftrightarrow p + e^-) + \lambda(n + e^+ \leftrightarrow p + \overline{\nu}_e) + \lambda(n \leftrightarrow p + e^- + \overline{\nu}_e).
\end{equation}
Assuming that all particle constituents share a common temperature sufficiently low to justify the use of the Boltzmann distribution instead of the Fermi-Dirac statistics, and neglecting the electron mass in comparison with the typical energies of electrons and neutrinos, the neutron abundance can be derived by examining the proton-to-neutron conversion processes. Within these approximations, straightforward analytical treatment yields an expression for the total transition rate, as presented in Refs.~\cite{Capozziello:2017bxm, Barrow:2020kug,Torres:1997sn,Lambiase:2011zz},
\begin{equation}
\lambda_{\text{tot}}(\mathcal{T}) = 8\mathcal{AT}^{3} \left( 12\mathcal{T}^{2} + 6\mathcal{QT} + \mathcal{Q}^{2} \right).
\end{equation}
Note that, the transition rates $\lambda_{np}(\mathcal{T})$ and $\lambda_{pn}(\mathcal{T})$ are connected through the relation 
$\lambda_{np}(\mathcal{T}) = e^{-\mathcal{Q}/\mathcal{T}}\lambda_{pn}(\mathcal{T})$. 
Here, $\mathcal{Q} = \tilde{m}_n - \tilde{m}_p = 1.29 \times 10^{-3}\,\mathrm{GeV}$ denotes the neutron-proton 
mass difference and $\mathcal{A} = 1.02 \times 10^{-11}\,\mathrm{GeV^{-4}}$ which is commonly interpreted as a constant parameter related to the amplitude of primordial density perturbations in the early Universe.
The freeze-out temperature corresponds to the following Hubble rate,
\begin{equation}\label{hubble at freeze out}
H(\mathcal{T}_f) = \lambda_{\text{tot}}(\mathcal{T}_f) \approx c_q\, \mathcal{T}_f^{5},
\end{equation}
where $c_q = 4\mathcal{A} \times 4! \approx 9.8 \times 10^{-10}\,\mathrm{GeV^{-4}}$.
Furthermore, $\lambda_{np}(\mathcal{T})$ can be represented as the cumulative contribution of the individual weak interaction processes described in equations \eqref{weak interaction}. Using Eq.
~\eqref{Hubble at temp} at $\mathcal{T} = \mathcal{T}_f$ and then \eqref{hubble at freeze out}, we obtain
\begin{equation}
\mathcal{T}_f = 
\left(
\frac{\pi^{2} g_*}{90 M_{p}^{2}} \, \mathcal{C}_{q}^{2}
\right)^{\!1/6}
\approx 0.6~\mathrm{MeV}.
\end{equation}

In modified gravity theories, additional terms can modify the Hubble parameter. For instance, in \( f(T,\mathcal{L}_m) \) gravity, the Hubble parameter is
\begin{equation}
 H = H_{\text{GR}} \left( 1 + \frac{\rho_{\text{DE}}}{\rho_r} \right)^{1/2},   
\end{equation}

where \( \rho_{\text{DE}} \) is the energy density due to dark energy. The deviation from the GR Hubble parameter is
\begin{equation}
\Delta H = H - H_{\text{GR}} = H_{\text{GR}} \left( \left( 1 + \frac{\rho_{\text{DE}}}{\rho_r} \right)^{1/2} - 1 \right) \approx \frac{\rho_{DE}}{\rho_r}\frac{H_{GR}}{2}.
\end{equation}

The above term inside the square root may be linearized to first order, since during the radiation-dominated epoch the dark energy density satisfies $\rho_{DE} \ll \rho_r$. 
Since \( H(\mathcal{T}_f) =c_q \mathcal{T}_f^5\), this implies \( \Delta H(\mathcal{T}_f)= 5c_q \mathcal{T}_f^4 \Delta \mathcal{T}_f\) and therefore this deviation leads to a shift in the freeze-out temperature,
\begin{equation}\label{deltaTf}
\frac{\Delta \mathcal{T}_f}{\mathcal{T}_f} \approx \frac{\rho_{\text{DE}}}{\rho_r} \frac{H_{\text{GR}}}{10 c_q \mathcal{T}_f^5}.
\end{equation}

To ensure consistency with observations, the deviation in the freeze-out temperature must satisfy,
\begin{equation}
\left| \frac{\Delta \mathcal{T}_f}{\mathcal{T}_f} \right| < 4.7 \times 10^{-4}.
\end{equation}

This constraint is derived from measurements of the primordial abundances of light elements, such as helium-4, deuterium, and lithium-7. The study of BBN in the framework of modified gravity theories provides valuable insights into the thermal history of the early Universe. By comparing theoretical predictions with observational constraints, one can place limits on the parameters of the alternative theories.

\section{The Model}\label{sec:model}
The functional form considered here is the following,
\begin{equation}\label{eq:hybmodel}
    f(T,{\mathcal{L}_m})=\frac{\alpha T_0^2}{T}+\beta {\mathcal{L}_m},
\end{equation}
where $\alpha$ and $\beta$ are model parameters without any physical dimension.

 We note that the TEGR term \(T\) is already included explicitly in the gravitational sector, and the function \(f(T,\mathcal{L}_m)\) acts only as an additional modification. Therefore, the choice \(f(T,\mathcal{L}_m)=\alpha T_0^2/T+\beta\,\mathcal{L}_m\) naturally leads to an effective structure of the form \(T + T^{-1}\). This guarantees a smooth recovery of TEGR in the high-torsion regime, while the inverse-torsion contribution becomes relevant only at late times, producing the desired deviations from standard cosmology.
 
 This class of model is considered because of its hybrid structure and its ability to accommodate both early- and late-time cosmological behavior \cite{Kolhatkar:2024oyy}. Its effectiveness in addressing cosmological tensions was demonstrated in \cite{Mishra:2025rhi}.
In the present work, we extend this framework by incorporating the matter Lagrangian ${\mathcal{L}_m}$. A related study, where the same model was analyzed in another matter coupling theory to explore cosmological tensions, can be found in \cite{Mishraptep}.

Now, substituting the model \eqref{eq:hybmodel} in equation \eqref{rho_{DE}}, we get the following form

\begin{equation}\label{rhode}
\rho_{\mathrm{DE}}
= 3 M_p^{2}
\left(
\frac{3 \alpha \, H_0^{4}}{H^{2}}
+ \frac{\beta\, \rho}{6}
\right).
\end{equation}

The second model parameter is reduced due to the dependency obtained using the condition $H(0)=H_0$ in the first motion equation. The relationship reads

\begin{equation}\label{betaparameter}
    \beta =-\frac{2 \left(9 \alpha  {H_0}^2-3 {H_0}^2+\frac{\rho_0}{M_p^2}\right)}{{\rho_0}}.
\end{equation}

During the radiation dominated era we have $\rho_m \ll \rho_r$, so that $\rho \approx \rho_r$. By substituting this into the dark energy density \eqref{rhode} and then inserting the resulting expression together with the energy density of relativistic particles \eqref{rho_r} and the relation \eqref{HGR} into \eqref{deltaTf}, we obtain

\begin{equation}\label{eq:dtf}
\frac{\Delta {\mathcal T_f}}{{\mathcal T_f}} = 
\frac{
g_*^{2}\,\pi^{4}\,\beta\, {\mathcal T_f}^{8}\, M_{p}
\;+\;
48600\, \alpha\, H_{0}^{4}\, M_{p}^{3}
}{
60 \sqrt{10}\; g_*^{3/2}\, \pi^{3}\, c_q\, {\mathcal T_f}^{11}
}.
\end{equation}

Substituting the model \eqref{eq:hybmodel} in the first Friedmann equation \eqref{first Friedmann} and using \eqref{betaparameter} gives rise to
the Hubble expression
\begin{equation}\label{eq:hub}
    H(z)=\sqrt{\frac{1}{2}\left(\sqrt{{H_0}^4 \left(12 \alpha +(1-3 \alpha )^2 (z+1)^6\right)}-(3 \alpha -1) {H_0}^2 (z+1)^3\right)}.
\end{equation}
The Eq.~\eqref{eq:dtf} will be used for early Universe analysis, while the Eq.~\eqref{eq:hub} will be applied to the late time constraints through observational datasets and statistical methodology.
\section{Observational Datasets}\label{sec:data}
To constrain the free parameters of the $f(T,\mathcal{L}_m)$ model, we employ a Bayesian statistical analysis based on the MCMC method. The likelihood functions are constructed from the observational datasets described in the following subsections, and the posterior probability distributions of the parameters are sampled using the affine-invariant ensemble sampler implemented in the \texttt{emcee} package \cite{Foreman-Mackey:2012any}. This algorithm is particularly well-suited for problems with correlated parameters, as it improves the efficiency of chain exploration in multi-dimensional parameter spaces.

Convergence of the chains is ensured by adopting multiple walkers with sufficiently long burn-in and sampling phases. The marginalized posterior distributions and confidence contours are obtained using the \texttt{GetDist} package \cite{lewis2019getdist}, which provides a robust framework for analyzing and visualizing MCMC chains. We report parameter constraints at the $68\%$ and $95\%$ confidence levels throughout this work. A more comprehensive discussion on the methodology can be found in \cite{Mishramnras}.
\subsection{Union3}
 The Union3 supernova sample is a homogeneous collection of 2087 type la 
 supernovae observations from 24 distinct datasets. It was developed by the Supernova Cosmology Project (SCP) \cite{Rubin:2023jdq}. 
 
 It updates and extends the earlier Union and Union2 datasets by incorporating more precise photometric calibrations, improved systematic corrections, and a larger uniformly reprocessed sample.
 Union3 contains over 1400 spectroscopically confirmed SNe Ia spanning a redshift interval of $z \sim 0.01$ to $z \sim 1.4$, providing crucial cosmological information for probing the expansion history of the Universe and constraining dark energy. It combines supernova observations from many major surveys (like SNLS, SDSS, Pan-STARRS, CSP, low-redshift programs, and HST), and all light curves are re-analyzed with the same SALT2 method to ensure uniformity. 
In the Union3 analysis, the Tripp formula provides the framework for standardizing Type Ia supernovae, allowing them to serve as reliable distance indicators, which is given as
\begin{equation}\label{Tripp}
    \mu = m_B - M + \alpha x_1 - \beta c ,
\end{equation}
where the observed peak magnitude ($m_B$) serves as the raw brightness measurement, while the light-curve stretch parameter ($x_1$) corrects for width luminosity correlations, and the color parameter ($c$) accounts for dust extinction effects. The absolute magnitude ($M$) provides the fundamental calibration for intrinsic luminosity, with scaling coefficients $\alpha$ and $\beta$ quantifying the strength 
of the stretch luminosity and color luminosity relationships. Through these components, the formula yields the distance modulus ($\mu$), transforming variable supernovae into precise cosmological distance indicators.
The covariance matrix, forms the basis for uncertainty quantification in Union3, ensuring that correlated errors are treated consistently across the full dataset, which is given as
\begin{equation}
    C = D_{\mathrm{stat}} + C_{\mathrm{sys}} ,
\end{equation}
 where $D_{\mathrm{stat}}$ represents the combined impact of measurement errors, photometric noise, and intrinsic dispersion, while $C_{\mathrm{sys}}$ accounts for systematic uncertainties arising from calibration, host galaxy effects, extinction, model assumptions, and peculiar motions. Their influence on the analysis was evaluated via Monte Carlo propagation within the SALT2 framework.
Standardization of SNe Ia as distance indicators uses the formula (\ref{Tripp}).
Cosmological constraints are then obtained by comparing observed distance moduli $\mu_{\mathrm{obs}}$ with theoretical predictions $\mu_{\mathrm{model}}(z;\theta)$ for a cosmological parameter set $\theta$, with the fit evaluated using the chi-squared statistic,
\begin{equation}
    \chi^2 = \mathcal{A}^{T}C^{-1}_{\mathrm{union}}\mathcal{A} ,
\end{equation}
where $\mathcal{A}=\big(\mu_{\mathrm{obs}} - \mu_{\mathrm{model}}\big)$. The dataset used in this work is summarized in \autoref{tab:Table 1}, and the covariance matrix $C_{\mathrm{union}}$ is provided in \cite{Rubin:2023jdq}.

\begin{table}
\centering
\caption{Data from Union3 compilation (Rubin et al., 2023)\cite{Rubin:2023jdq}.}
\begin{tabular}{lcr}
\hline
$z$ & $\mu(z)$ & $\mu_{\text{err}}$ \\
\hline
0.050000 & 36.630361 & 0.0086044 \\
0.100000 & 38.235499 & 0.0078677 \\
0.150000 & 39.147691 & 0.0085239 \\
0.200000 & 39.834880 & 0.0082393 \\
0.250000 & 40.340319 & 0.0083980 \\
0.300000 & 40.813875 & 0.0083796 \\
0.350000 & 41.202487 & 0.0084498 \\
0.400000 & 41.568946 & 0.0085916 \\
0.450000 & 41.822650 & 0.0087093 \\
0.500000 & 42.136311 & 0.0087613 \\
0.550000 & 42.352181 & 0.0089000 \\
0.600000 & 42.599505 & 0.0090432 \\
0.650000 & 42.793121 & 0.0093150 \\
0.700000 & 42.945322 & 0.0095681 \\
0.750000 & 43.125351 & 0.0095079 \\
0.800000 & 43.319655 & 0.0096321 \\
0.898138 & 43.647594 & 0.0109719 \\
0.996276 & 44.011256 & 0.0110380 \\
1.094415 & 44.191092 & 0.0150685 \\
1.232244 & 44.577985 & 0.0138742 \\
1.390961 & 44.842358 & 0.0125267 \\
2.262260 & 45.997159 & 0.1155183 \\
\hline
\end{tabular}

\label{tab:Table 1}
\end{table}

\begin{eqnarray}\label{C_union}
    \centering
C_{\text{union}} =
\begin{bmatrix}
0.0086044 & 0.0078396 & \cdots & 0.0081695 \\
0.0078396 & 0.0078677 & \cdots & 0.0079259 \\
\vdots    & \vdots    & \ddots & \vdots    \\
0.0081695 & \cdots    & \cdots & 0.1155183
\end{bmatrix}_{22 \times 22}
\end{eqnarray}

\subsection{Cosmic Chronometers}  
Understanding the expansion history of the Universe is a key goal of modern cosmology. Throughout cosmic history, the rate of universe expansion has varied: it initially slowed due to the mutual gravitational attraction of all the matter present, and in more recent epochs, it has accelerated, a process attributed to the influence of ``dark energy". The scale factor $a(t)$ describes how the universe expands with cosmic time. For a given cosmological model that specifies the contributions of different energy components, the functional form of $a(t)$ is uniquely determined.

Since the exact composition of the universe is not fully understood, $a(t)$ must be constrained through observations. Its evolution is connected to the Hubble parameter.
To determine $H$ as a function of cosmic time, various observational methods have been suggested, such as ``standard candles'' (like Type Ia supernovae) and ``standard rulers'' (for instance, Baryon Acoustic Oscillations). An alternative independent method is offered by the differential dating of ``cosmic chronometers", a technique first introduced by Jimenez and Loeb (2002)\cite{Jimenez:2001gg}. This approach enables the calculation of the expansion rate without requiring consideration of the specific metric between the chronometer and us, providing a conceptually straightforward method that avoids assumptions related to integrated observables.

\par

In terms of redshift, the expansion rate is defined as:
\begin{equation}
    H(z)=-\frac{1}{1+z}\frac{dz}{dt},
\end{equation}
where $\frac{dz}{dt}$ can be deduce from $\frac{\Delta z}{\Delta t}$. Here, $\Delta z$ signifies the variation in redshift, while $\Delta t$ denotes the variation in age between two galaxies. Since the redshift $z$ can be measured with high accuracy (e.g., spectroscopic redshifts of galaxies have typical uncertainties $\sigma_{z}\leq 0.001$), this relation allows for a direct estimate of $H(z)$ using a differential approach. Unlike absolute age determinations, which are subject to large systematic uncertainties, this method relies on relative age differences between galaxies at nearby redshifts, thereby significantly reducing modelling errors.

 This investigation employs a dataset comprising 34 cosmic chronometer (CC) measurements of the Hubble parameter, aggregated from the works of these references \cite{Moresco:2012jh, Moresco:2016mzx, Jimenez:2003iv}. These measurements span the redshift interval from approximately $z \sim 0.1$ to $z \sim 2$ and are of particular importance due to their acquisition through cosmology-independent methodologies. Consequently, they serve as a robust means for validating results and providing supplementary constraints when integrated with other cosmological probes, such as Type Ia supernovae, baryon acoustic oscillations, and gravitational wave standard sirens.

The total chi-square contribution from the CC dataset is evaluated by combining the terms from both the uncorrelated and correlated measurements, namely
\begin{equation}
\chi^2_{\text{CC}} = \chi^2_{\text{uncorr}} + \chi^2_{\text{corr}}.
\end{equation}
The uncorrelated part, consisting of 19 data points, is computed as
\begin{equation}
\chi^2_{\text{uncorr}} = 
\sum_{i=1}^{19} 
\frac{\left[ H_{\text{th}}(z_i) - H_{\text{obs}}(z_i) \right]^2}
{\sigma_H^2(z_i)},
\end{equation}
whereas the correlated contribution from the remaining 15 points is given by
\begin{equation}
\chi^2_{\text{corr}} = 
\Delta H^{T} C^{-1} \Delta H,
\end{equation}
with $\Delta H = H_{\text{th}} - H_{\text{obs}}$ denoting the vector of residuals and $C$ the $15 \times 15$ covariance matrix associated with the correlated CC measurements~\cite{Moresco:2020fbm}.

\subsection{Pantheon+SH0ES (SN22)}
Throughout this work, we refer to this data as SN22. The Pantheon+SH0ES supernova sample stands out as one of the most accurate resources for studying the recent expansion of the universe and measuring the local Hubble constant, $H_0$. This sample utilizes a hierarchical distance ladder approach, beginning with Cepheid variables for nearby calibrations and extending to Type Ia supernovae (SNe Ia) for greater distances \cite{Scolnic:2021amr, Brout:2022a, Brout:2022b}. The dataset combines detailed observations of Cepheids with a comprehensive collection of 1,701 spectroscopically confirmed SNe Ia from 1,550 host galaxies, sourced from over 50 different observational surveys. It covers a broad redshift range from as low as 0.001 up to about 2.3 \cite{Scolnic:2021amr, Brout:2022a}. Notably, 77 of these supernovae occur in galaxies that also contain Cepheid variables. This overlap enables a direct calibration of the absolute luminosities of SNe Ia, which is crucial for disentangling the relationship between the Hubble constant and the intrinsic brightness of these supernovae \cite{Riess:2022a, Brout:2022a}.

Each SN Ia light curve is processed using the SALT2 light-curve fitting model and standardized with the Tripp relation to account for intrinsic variations in SN luminosities \cite{Scolnic:2021amr}. The resulting distance moduli include a full covariance matrix that incorporates both statistical and systematic uncertainties \cite{Brout:2022a}. The chi-squared statistic is defined as
\begin{equation}
\chi^2_{\text{SN}} = \Delta \mu^T C_{\text{SN}}^{-1} \Delta \mu,
\end{equation}
where $\Delta \mu$ represents the difference between observed and theoretical distance moduli, and $C_{\text{SN}}$ denotes the covariance matrix \cite{Brout:2022a, Perivolaropoulos:2023a}.

The theoretical distance modulus is expressed as
\begin{equation}
\mu_{\text{th}}(z) = 5 \log_{10}\left(\frac{D_L(z)}{1\,\text{Mpc}}\right) + 25,
\end{equation}
where the luminosity distance $D_L(z)$ is given by
\begin{equation}
D_L(z) = c (1 + z) \int_0^z \frac{dx}{H(x; \theta)}.
\end{equation}
Here, $H(x; \theta)$ is the Hubble parameter for the chosen cosmological model and $\theta$ represents the set of model parameters \cite{Scolnic:2021amr}.

For Pantheon+, the residual vector is
\begin{equation}
\Delta \mu_i = \mu_{\text{obs},i} - \mu_{\text{th}}(z_i).
\end{equation}
In the Pantheon+SH0ES analysis, this is modified to account for Cepheid calibration:
\begin{equation}
\tilde{\Delta \mu}_i =
\begin{cases}
\mu_{\text{obs},i} - \mu^{\text{Ceph}}_i, & \text{for Cepheid-hosting galaxies},\\
\mu_{\text{obs},i} - \mu_{\text{th}}(z_i), & \text{otherwise}.
\end{cases}
\end{equation}
This approach anchors the SN luminosity scale and provides a model-independent calibration of $H_0$ \cite{Riess:2022a, Brout:2022a}.

The Pantheon+SH0ES dataset has emerged as a key resource in contemporary cosmology and is frequently combined with other observational probes, such as cosmic chronometers, which offer direct measurements of the Hubble parameter across different redshifts, and baryon acoustic oscillations, which provide a standard ruler for large-scale structure. This combination allows for precise constraints on cosmological parameters, enabling rigorous tests of both the standard \(\Lambda\)CDM model and various extended or alternative cosmological frameworks \cite{Brout:2022a, Perivolaropoulos:2023a}
 
\section{Results}\label{sec:results}
In this section, we present a comprehensive discussion of the cosmological
implications of the proposed $f(T,\mathcal{L}_m)$ model across different
epochs of the Universe. 

\begin{figure}
    \centering
    \includegraphics[width=0.65\linewidth]{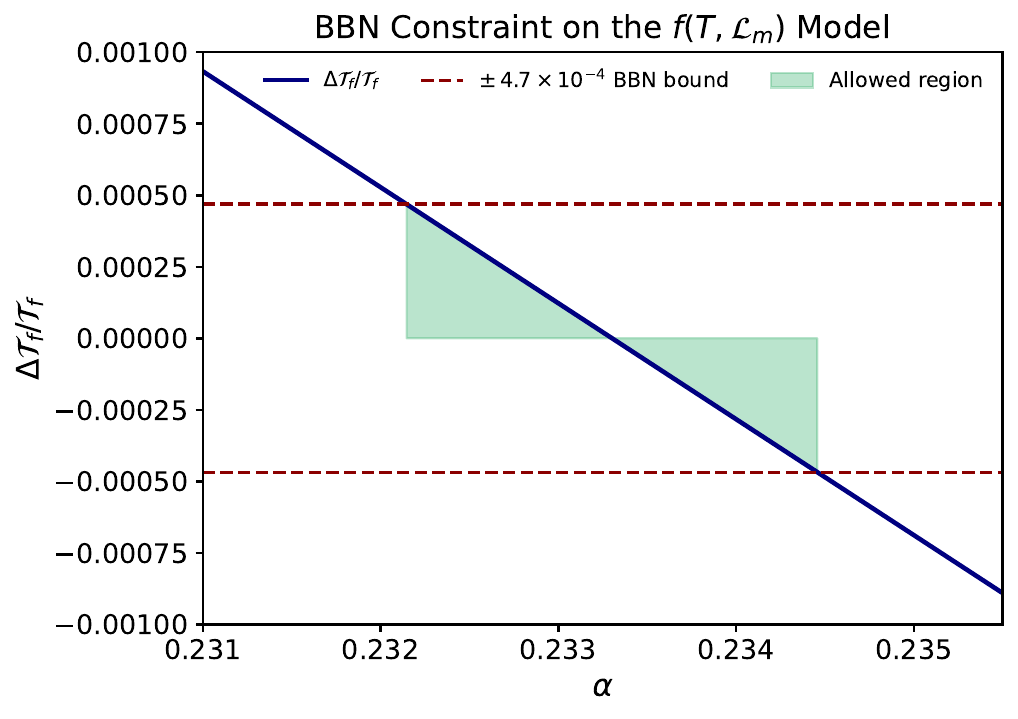}
    \caption{BBN bound on $\alpha$. The curve shows 
$\Delta\mathcal{T}_f/\mathcal{T}_f$, with dashed lines marking the limit 
$|\Delta\mathcal{T}_f/\mathcal{T}_f| < 4.7\times10^{-4}$. The shaded region
indicates the allowed values of $\alpha$.}

    \label{fig:bbn}
\end{figure}

To incorporate early-Universe constraints, we apply the BBN bound
$|\Delta\mathcal{T}_f/\mathcal{T}_f| < 4.7\times10^{-4}$ to the modified 
freeze-out temperature expression (Eq.~\eqref{eq:dtf}). 
This yields a narrow allowed interval for the inverse-torsion parameter,
$0.2321 \lesssim \alpha \lesssim 0.2346$. 
Figure~\ref{fig:bbn} summarizes this bound, where the dashed horizontal lines
represent the BBN limit and the shaded region indicates the viable range of
$\alpha$.

\begin{figure}
    \centering
    \includegraphics[width=0.65\linewidth]{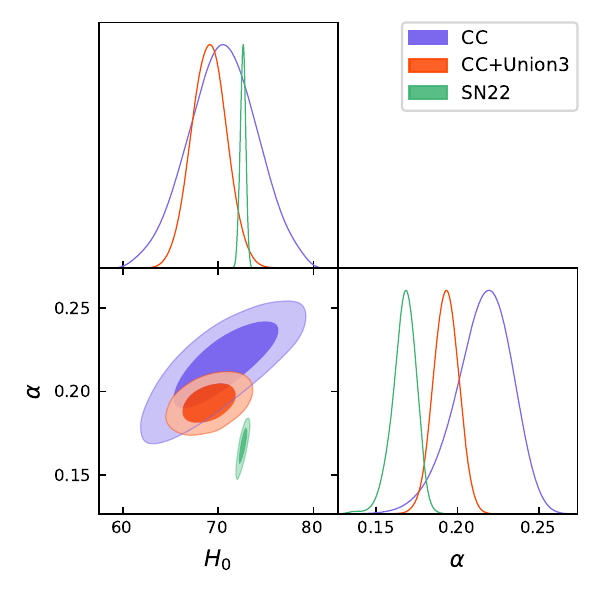}
    \caption{The 2D contours upto $2\sigma$ for the parameter space $\{H_0,\alpha\}$.}
    \label{fig:con}
\end{figure}

Based on the MCMC analysis with the observational datasets introduced in the
previous section, the reduced parameter space $\{H_0, \alpha\}$ is constrained
for three dataset combinations: CC, CC+Union3, and SN22. The contour
plots in \autoref{fig:con} display the confidence regions up to the $2\sigma$
level, with darker and lighter shades corresponding to the $1\sigma$ ($68\%$
CL) and $2\sigma$ ($95\%$ CL) limits, respectively. The inferred $1\sigma$
constraints are listed in \autoref{tab:Table 2}, showing that the
preferred values of $\alpha$ lie close to the BBN-allowed interval for all
three dataset combinations.

\begin{figure}
    \centering
    \includegraphics[width=0.49\linewidth]{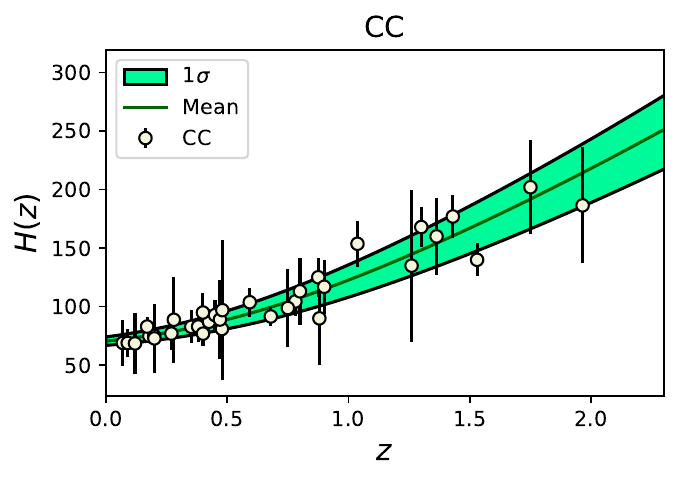}
    \includegraphics[width=0.49\linewidth]{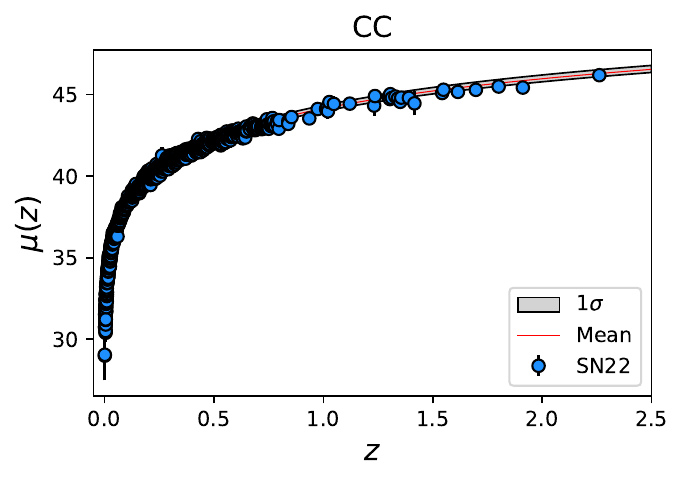}
    \includegraphics[width=0.49\linewidth]{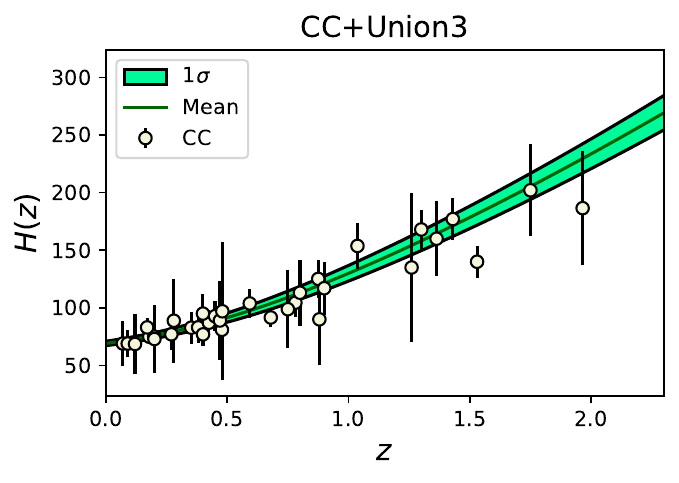}
    \includegraphics[width=0.49\linewidth]{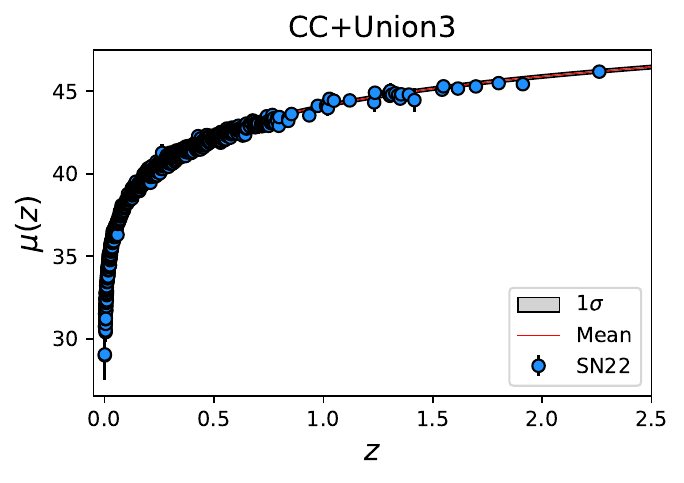}
    \includegraphics[width=0.49\linewidth]{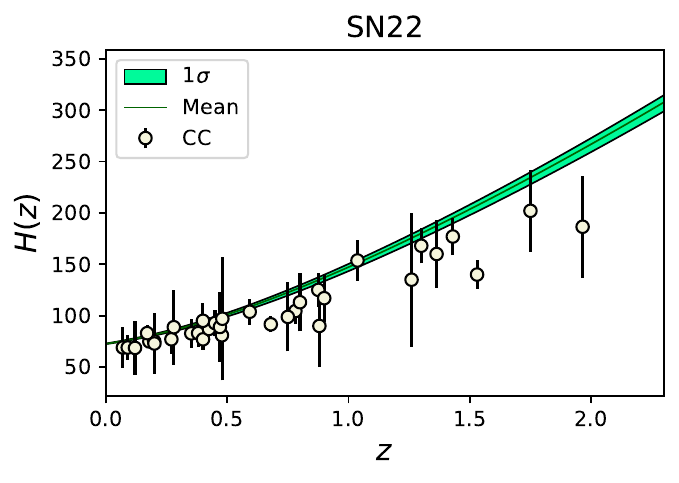}
    \includegraphics[width=0.49\linewidth]{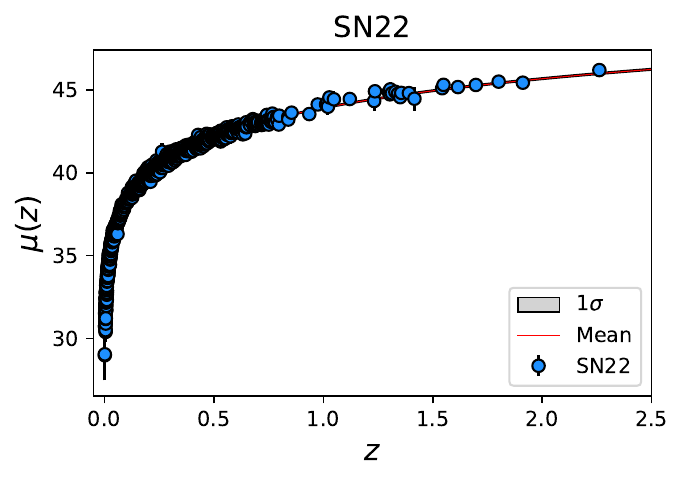}
    \caption{The left panel presents Hubble parameter evolution against the 34 CC data, and the right panel presents the 1701 SN22 data.}
    \label{fig:hmu}
\end{figure}

\begin{figure}
    \centering
    \includegraphics[width=0.49\linewidth]{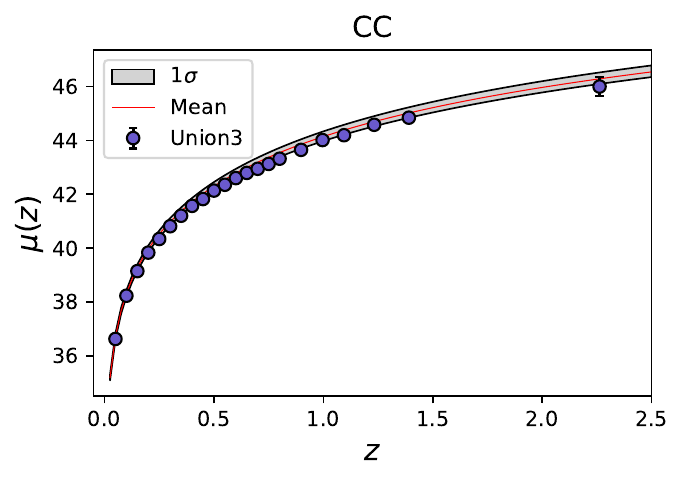}
    \includegraphics[width=0.49\linewidth]{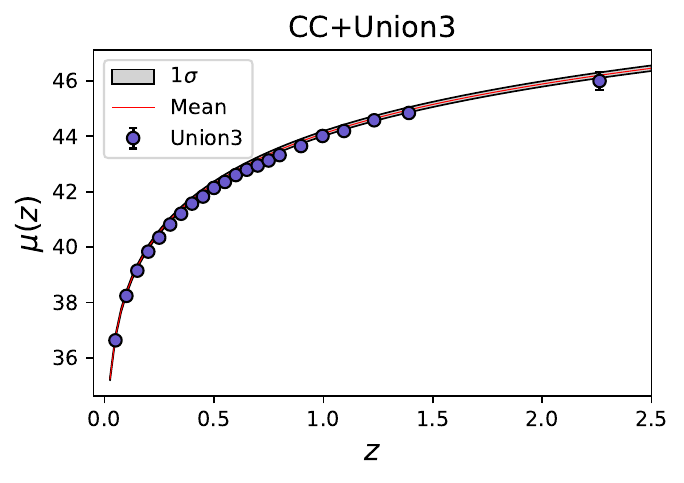}
    \includegraphics[width=0.49\linewidth]{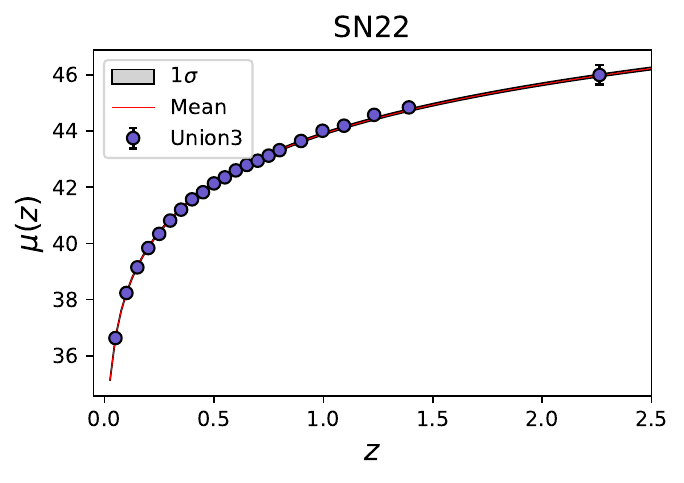}
    \caption{Distance modulus profile against redshift and Union3 datapoints.}
    \label{fig:muni}
\end{figure}

Using the best-fit values and their uncertainties, we reconstruct the mean and
$1\sigma$ confidence regions for key cosmological quantities: the Hubble
parameter, distance modulus, deceleration parameter, and the effective
equation-of-state (EoS) parameter over the redshift range $z\in[0,2.5]$.
\autoref{fig:hmu} (left) shows the reconstructed Hubble parameter $H(z)$
together with the 34 CC data points. The corresponding distance modulus
against the 1701 SN22 data points is shown in the right panel, displaying
excellent agreement with the observations. In addition,
\autoref{fig:muni} confirms that the reconstructed distance modulus remains in
good agreement with the Union3 dataset across the full redshift range.

\begin{figure}
    \centering
    \includegraphics[width=0.49\linewidth]{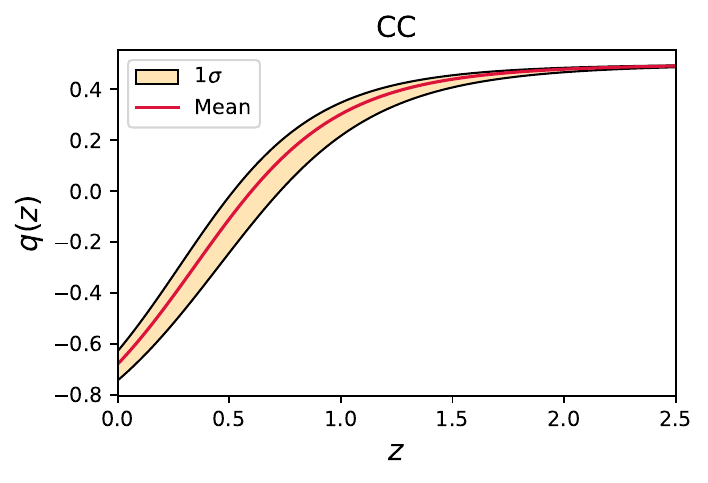}
    \includegraphics[width=0.49\linewidth]{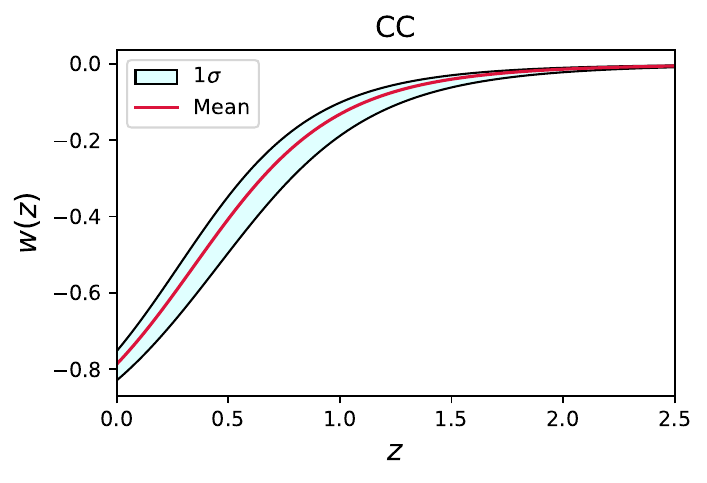}
    \includegraphics[width=0.49\linewidth]{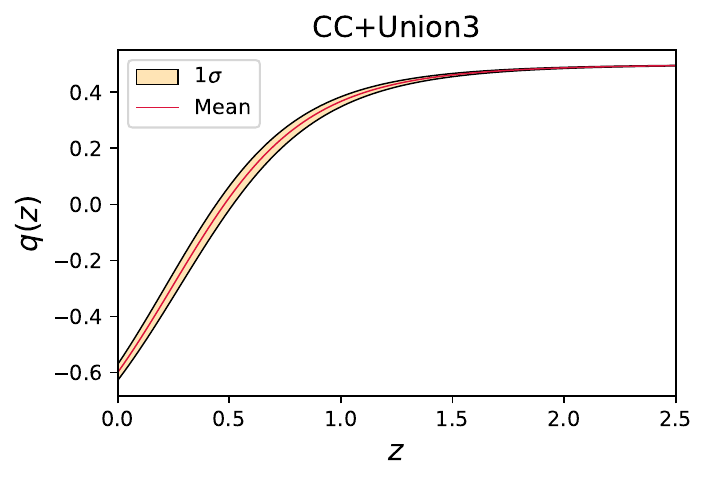} 
    \includegraphics[width=0.49\linewidth]{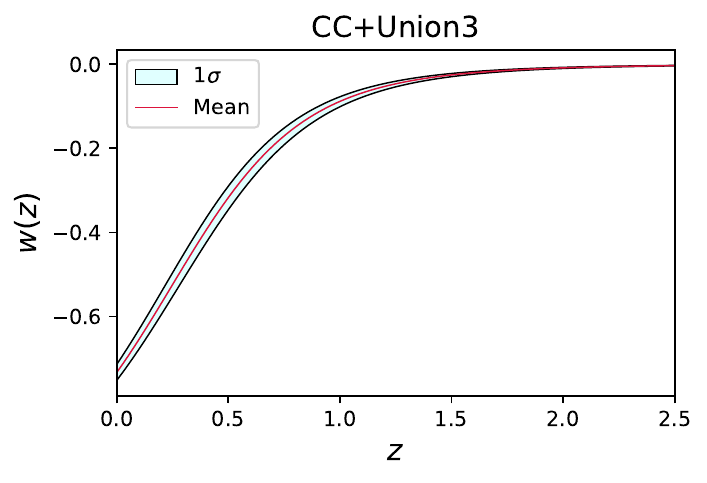}
    \includegraphics[width=0.49\linewidth]{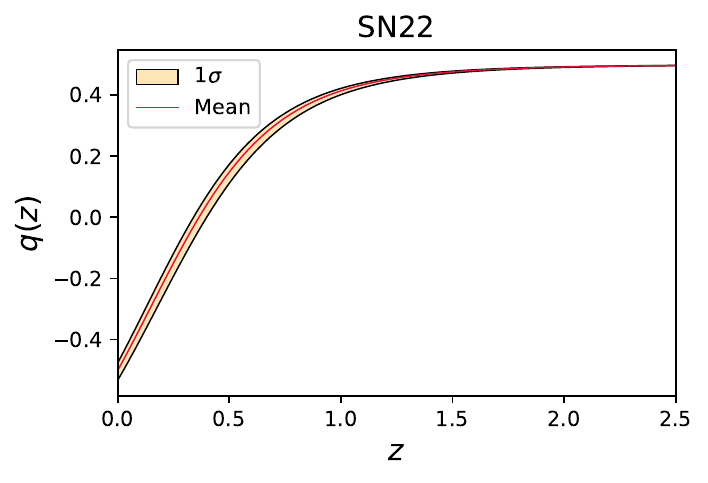}
    \includegraphics[width=0.49\linewidth]{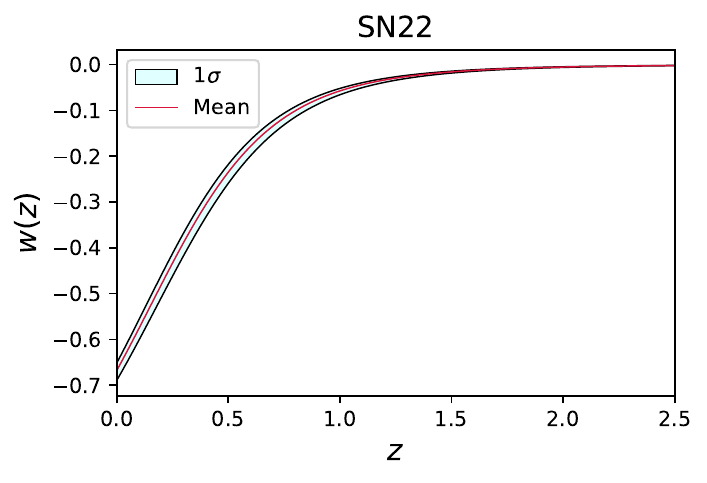}
    \caption{Evolution of the Universe through deceleration parameter (left panel) and EoS parameter (right panel.)}
    \label{fig:qw}
\end{figure}

The reconstructed deceleration parameter $q(z)$ and effective EoS parameter
$w(z)$ are presented in \autoref{fig:qw}. The evolution of $q(z)$ shows a
smooth transition from a decelerated phase ($q>0$) at high redshifts to an
accelerated phase ($q<0$) at low redshifts. The transition redshift $z_t$,
where $q(z)$ crosses zero, is well constrained and reported in
\autoref{tab:Table 3}. The obtained values of $z_t$ are consistent with
the kinematic constraints inferred from SNe Ia and $H(z)$ data in
Refs.~\cite{Cunha:2008mt, Farooq:2013hq, Capozziello:2015rda, Rani:2015lia},
and with expectations from the standard cosmological model.

The evolution of $w(z)$ remains negative over the full redshift interval,
indicating the dominance of an effective dark-energy-like component driving the
present cosmic acceleration. The present-day value $w_0$ is consistently found
to satisfy $w_0 > -1$ across all dataset combinations (see
\autoref{tab:Table 3}), suggesting a quintessence-like behaviour of the
effective dark energy in the $f(T,\mathcal{L}_m)$ framework rather than a
cosmological constant ($w=-1$) or a phantom regime ($w<-1$).
\begin{table}
\centering
\caption{$1\sigma$ results obtained from the MCMC for each combination.}
\begin{tabular}{lcc}
\hline
Dataset & $H_0$ & $\alpha$ \\
\hline\\
CC & ${70.5}\pm{3.6}$ & ${0.216}_{-0.015}^{+0.020}$\\\\
CC+Union3 & ${69.1}\pm{1.9}$ & ${0.1929}\pm{0.0079}$ \\\\
 SN22 & ${72.61}_{-0.28}^{+0.31}$ & ${0.1671}_{-0.0061}^{+0.0082}$ \\\\
\hline
\end{tabular}

\label{tab:Table 2}
\end{table}

\begin{table}
\centering
\caption{$1 \sigma$ values for the present time deceleration parameter ($q_0$), EoS parameter ($w_0$), and the transition redshift ($z_t$).}
\begin{tabular}{lccc}
\hline
Dataset & $q_0$ & $w_0$ & $z_t$ \\
\hline\\
CC & ${-0.67}_{-0.07}^{+0.05}$ & ${-0.78}_{-0.04}^{+0.03}$ & ${0.59}_{-0.08}^{+0.13}$  \\\\
CC+Union3 & ${-0.59}_{-0.03}^{+0.02}$ & ${-0.73}_{-0.02}^{+0.02}$ &${0.47}_{-0.03}^{+0.04}$  \\\\
 SN22 & ${-0.50}_{-0.03}^{+0.03}$ & ${-0.66}_{-0.02}^{+0.01}$ & ${0.36}_{-0.03}^{+0.03}$ \\\\
\hline
\end{tabular}

\label{tab:Table 3}
\end{table}


\section{Concluding Remarks}\label{sec:conclusions}
In this work, we have carried out a systematic investigation of the cosmic
expansion history within the framework of $f(T,\mathcal{L}_m)$ gravity. The
model parameters were constrained using increasingly comprehensive datasets,
starting from cosmic chronometers and extending to the Union3 and SN22
supernova compilations, leading to progressively tighter bounds on the reduced
parameter space. The reconstructed Hubble and distance-modulus functions show
excellent consistency with these observations, supporting the phenomenological
viability of the $f(T,\mathcal{L}_m)$ scenario. In addition, the inclusion of
the BBN bound further restricts the inverse-torsion
parameter $\alpha$ to a narrow range, ensuring compatibility with early-time
physics.

A key result of our analysis is the emergence of a quintessence-like behaviour
with $w_0 > -1$, pointing toward a dynamical origin of late-time cosmic
acceleration in this framework. The model also predicts an earlier transition
to the accelerated phase compared to $\Lambda$CDM, indicating subtle
deviations in the background evolution that may be relevant to late-time
cosmological tensions. These findings constitute the first detailed
reconstruction of the background dynamics in $f(T,\mathcal{L}_m)$ gravity with
both early- and late-time constraints.

Future extensions of this work may include large-scale structure, weak lensing,
and CMB observations to test the robustness of the model across multiple
cosmological probes. Overall, our results suggest that $f(T,\mathcal{L}_m)$
gravity offers a consistent and flexible alternative framework for describing
cosmic acceleration while remaining compatible with current observational data.

\section*{Data availability} No new data were generated or analysed in support of this research.

\section*{Acknowledgement} SSM acknowledges the Council of Scientific and Industrial Research (CSIR), Govt. of India for awarding Senior Research fellowship (E-Certificate No.: JUN21C05815). SP acknowledges the Council of Scientific and Industrial Research (CSIR), Govt. of India, for providing Junior Research Fellowship (E-certificate No.: 24D/01/00653). PKS acknowledges Anusandhan National Research Foundation (ANRF), Department of Science and Technology (DST), Government of India for financial support to carry out Research project No.: CRG/2022/001847 and IUCAA, Pune, India for providing support through the visiting Associateship program. We are very grateful to the honorable referee and to the editor for the illuminating suggestions that have significantly improved our work in terms of research quality and presentation.

\bibliography{main}

\end{document}